\documentclass[scrartcl,onecolumn,preprintnumbers,amsmath,amssymb,aps,prb,12pt]{revtex4}
\renewcommand{\baselinestretch}{2}
\usepackage{graphicx}
\usepackage{color}
\usepackage{dcolumn}
\usepackage{bm}
\usepackage{listings}
\usepackage[version=3]{mhchem}
\usepackage{etoolbox}
\usepackage{marvosym}
\usepackage[pdfborderstyle={/S/U/W 0}]{hyperref}

\usepackage{url}
\usepackage{ifpdf}

\begin{document}
\pagestyle{headings}
\pagenumbering{arabic}
\pdfinfo{
/Author (Andreas Lubatsch; Regine Frank)
/Title (Tuning the Quantum Efficiency of Random Lasers - Intrinsic
Stokes-Shift and Gain)
/Keywords (random laser, transport of light, Vollhardt, W\"olfle, Cooperon,
nonlinearity, Bethe Salpeter Equation)
}

\title{Tuning the Quantum Efficiency of Random Lasers - Intrinsic Stokes-Shift and Gain}
\author{Andreas Lubatsch$^{1,2}$ and Regine Frank$^{3,4}$}
\email[Correspondence should be addressed to:]{regine.frank@googlemail.com}
\affiliation{$^1$ Electrical Engineering, Precision Engineering, Information
  Technology, Georg-Simon-Ohm University of Applied Sciences, Kesslerplatz 12,
  90489 N\"urnberg, Germany\\
$^2$ Physikalisches Institut, Rheinische Friedrich-Wilhelms Universit\"at
Bonn, Wegelerstr. 8, 53115 Bonn, Germany\\
$^3$ Institute of Theoretical Physics, Optics and Photonics, Eberhard-Karls-Universit\"at T\"ubingen, Auf der Morgenstelle 14, 72076 T\"ubingen, Germany\\
$^4$ Institute of Solid State Physics, Karlsruhe Institute of Technology
(KIT), Wolfgang-Gaede Strasse 1, 76131 Karlsruhe, Germany}

\begin{abstract}
\noindent 
\renewcommand{\baselinestretch}{2}
We report the theoretical analysis for tuning the quantum efficiency of solid
state random lasers. Vollhardt-W\"olfle theory of photonic
transport in disordered non-conserving and open random media, is coupled to lasing dynamics and solved positionally
dependent. The interplay of non-linearity and homogeneous non-radiative
frequency conversion by means of a Stokes-shift leads to a reduction of the quantum efficiency of
the random laser. At the threshold a strong decrease of the
spot-size in the stationary state is found due to the increase of
non-radiative losses. The coherently emitted photon number per unit of modal
surface is also strongly
reduced. This result allows for the conclusion that
Stokes-shifts are not sufficient to explain confined and extended mode regimes.
\end{abstract}
\maketitle

\section*{Introduction}
\label{introduction}
\renewcommand{\baselinestretch}{2}
The first description of stimulated emission was given by Albert Einstein in
1916 following the consideration of Max Planck who connected a
radiation field with the resonator mode. 44 years later Theodore Maiman presented the first experimental realization
of a laser. Random lasers have been predicted in 1967 by Lethokov
{\hyperlink{Let1}{\cite{Lethokov}}}$^,${\hyperlink{Let2}{\cite{Lethokov2}}}. They work in principle according to the same rules as the
conventional laser, only one significant difference exists: The cavity on first
sight is missing {\hyperlink{Cao98}{\cite{Cao98}}}.
Solid state random lasers consist of strongly scattering material which provides amplification on the basis of
multiple scattering procedures. The amplifier consists of granular matter {\hyperlink{MujumdarMie}{\cite{MujumdarMie}}} or a
perforated amplifier, where the scattering
strength is tuned with the density. The procedure is a random walk of the photon which is able to
undergo interference effects if the coherence is not violated too much by
inelastic scattering processes. However it does not exclude that
incoherent light propagating through the sample additionally or even majorly
inverts the electronic system of the material. Taking for granted that so
called closed loops of traveling and interfering photons may occur, it is
important to note that these effects are for sure not sufficiently strong to drive the system into lasing. They rather provide a large scale
trigger which leads to stimulated emission on a well defined
lasing area {\hyperlink{Frank14}{\cite{Frank14}}}. Hence the random laser could be a large scale single-mode laser if
not local decoherence effects would actually detune it {\hyperlink{Mujumdar04}{\cite{Mujumdar04}}}. Local in this sense
means that the local production of phonons at a certain position influences
the energy conservation and naturally leads to a spectral broadening of the
propagating photonic density. In combination with deviations in the local gain
spectrum a multi-modal regime is unavoidable and experimentally
observed. It is a challenge to tune the modal regime and select with respect
to size, position and frequency several modes, and to deplete others {\hyperlink{Moskmode}{\cite{Moskmode}}}. A
possible co-existence of extended and confined mode {\hyperlink{Kalt}{\cite{Kalt}}}, meaning a spatial
overlap consequently means that these modes must be energetically well
separated due to their gain spectrum. The question we are answering in this
article is, can we tune by temperature effects, the use of a Stokes-shift
during light-matter interaction, a transition between two modal
regimes, can phonons be used not just to tune the laser's frequency but also to
shape the mode and how flexible is this tuning? This ansatz is not to be
compared with a vibrational detuning of mesoscopic transport processes
{\hyperlink{Heat}{\cite{Heat}}}. Here we focus on the electronic subsystem of the
    solid. The electronic transition
probabilities are Stokes-detuned, which means we use a broadening of the photonic
spectrum due to phonon production. Stokes-shifted photons have
naturally a different mean free path and feature another absorption spectrum
themselves. Such conversion increases the laser threshold of the original
mode. When the number of converted photons is large, this mechanism
leads to separate thresholds and so called distributed feedback lasers (DFB).

We focus our considerations here to a three dimensional thin
sample that is displayed in {\hyperlink{exp}{Fig. \ref{exp}}}. The sample is unbounded
in-plane, meaning, it shall be very large compared to every other length scale
in the random laser. Shown is a possible sketch of a monodisperse densly
packed sample of zinc white (ZnO) spheres with a diameter of $d=260\,
nm$. These spheres have
bi-functional quality. They scatter light in the mesoscopic sense, but even
more they absorb and amplify photonic intensity and act as the semi-conductor laser material. So we find two accumulation processes:
First the mesoscopic accumulation which is enhanced for an increased particle
density and second a retardation effect through an enhanced life-time of
the electron-photon coupling in the gain region of the electronic
bandstructure. This property is special for pure ZnO solid state random
lasers. The density of our sample is assumed for the here presented
calculations to be of $50\%$ volume filling fraction which
corresponds to experimentally accessible values {\hyperlink{Kalt}{\cite{Kalt}}}. Due to the high
filling the scattering mean free path $l_s$ of photons is comparably short and
of the order of the wavelength of the scattered light. The transport
velocity $v_t$ hence is drastically reduced by the high number of occurring
scattering events. It has to be emphasized that mesoscopic processes act as
accumulation mechanism of photon density whereas the lasing frequency is
dominated by the materials absorption- and emission spectrum. The
semi-conductor bandstructure however may change due to intense pumping processes.\\
If we allow a Stokes-conversion as loss mechanism within the medium, it has to
be clarified how the spot sizes of random laser modes react on this
losses. The spot size is intrinsically connected to the degree of photonic
correlation on the one hand side as well as to the amount of coherently
emitted lasing intensity on the other hand. Our results will give
insight to the relation between spot size and coherent intensity which
will question the interpretation of the lasing spot as a {\it cavity}. This
leads ultimately to the question whether a co-existing modal regime is
achievable by the Stokes-tuning or whether a transition between confined and
extended modes is in principle unreachable by tuning the material
e.g. chemically and homogeneously in space.

\section*{Laser Dynamics - Tuning via Stokes-Shift}
\label{laser}
\renewcommand{\baselinestretch}{2}
In this work we consider ZnO nano-spheres providing the scattering and the
gain-channels for random lasing action. The impinging pump laser light
separates electron and atomic core within the
semi-conductor lattice structure of the bulk of the
pumped nano-sphere. Excitonic states are created which melt for high
excitation power due to short laser pulses to an electron hole
plasma. We consider here quasistationary pumping of $1.8 MW/cm^2$ to $2.4
MW/cm^2$. 
To describe lasing action, the electronic dynamics has to be accounted for
{\hyperlink{FlorescuMaster}{\cite{FlorescuMaster}}}$^,${\hyperlink{John04}{\cite{John04}}}. For the atomic level laser rate system
{\hyperlink{Siegman}{\cite{Siegman}}} consisting of four coupled energy levels (see
{\hyperlink{LRG}{Fig. \ref{LRG}}}) we write
\begin{eqnarray}
\frac{d}{dt} n_3(t)&=& \gamma_P -(1/\tau_{32})n_3\label{lrg4}\nonumber\\
\frac{d}{dt}  n_2(t) &=& (1/\tau_{32})n_3(t)- (1/\tau_{sp}) n_2(t)\label{lrg2}\nonumber\\
 &-&(1/\tau_{21})[n_2(t) - n_1(t)]n_{Ph}(t) - (1/\tau_{nr}) n_2(t)\nonumber\\
\frac{d}{dt}  n_1(t) &=& -(1/\tau_{10})n_1(t) + (1/\tau_{sp}) n_2(t)\label{lrg3}\nonumber\\
 &+&(1/\tau_{21})[n_2(t) - n_1(t)]n_{Ph}(t) + (1/\tau_{nr}) n_2(t)\nonumber\\
\frac{d}{dt}  n_0(t)&=&(1/\tau_{10})n_1(t) - \gamma_p\label{lrg4}
\label{lrgtot}
\end{eqnarray}

In the preceeding equations Eqs.(\ref{lrgtot}), $\gamma_P$ is the external pump rate
for two photon pumping, $n_{0-3}$ are electronic populations of the levels respectively,
$\tau_{ij}$ are the states' lifetimes $\frac{1}{\tau_{ij}}=\gamma_{ij}$,
 $\tau_{sp}$ represents the spontaneous
decay time and $\tau_{21}$ is the time scale of the lasing
transition. $\tau_{nr}$ is the non-radiative decay time which denotes the
investigated transition in order to tune the random laser via Stokes-shift.
The term $[n_2(t) - n_1(t)]n_{Ph}(t)$ marks the inversion of the occupation
numbers of level $1$ and $2$ proportional to the number of stimulated
emitted photons $n_{ph}$. 
All spatial coordinates are suppressed in Eqs.(\ref{lrgtot}) for the clarity of
presentation, however is noted that the photon numbers differ according to the position within the sample. The produced photon number $n_{ph}$ is to be read as
stimulated emission per excited ZnO atom. The material is considered with a
molar mass of $81,39 g\cdot mol^{-1}$ and the density of $5,61 g\cdot cm^{-3}$.
The Stokes-shift frequency-converts a certain fraction of photons, whereas the
loss (Stokes) or gain (anti-Stokes) of energy leads to a red or blue shift and so a
broadening of the spontaneous emission spectra of ZnO. It has to be noted that
in ZnO usually the Stokes shift is too small to generate another lasing
mode. Also significant heating or cooling is expected to occur
    with a strong
Stokes-shift. This however is not observed in ZnO but the frequency broadening is measured. Therefore the frequency-converted number of photons are simply
considered as loss for the amplified mode.  Consequently the non-linear
feedback mechanism is working less efficiently and the threshold of the laser is
increasing.\\
We will see in the following that not only the electronic procedures feel these losses,
additionally interference effects described diagrammatically by maximally crossed
diagrams (Cooperons) are reduced due to incoherent scattering. All
transitions in the above described system are not independent of each
other. The loss is intimately connected with the number of excited atoms and
consequently with the pump intensity. This leads to the assumption that gain
may compensate loss at some point, however the coherence properties of the
resulting mode are fundamentally different than those of a mode in a passive,
energy conserving medium.

\section*{Self-Consistent Transport Theory of Photons}
\label{Bethe-Salpeter}
\renewcommand{\baselinestretch}{2}
In preceeding work it has been shown that diagrammatic transport
{\hyperlink{Frank06}{\cite{Frank06}}}$^,${\hyperlink{Anderson}{\cite{Anderson}}} gives precise results for diffusive and localizing
photons in complex random media. We constitute a diagrammatic field theory ansatz for light in a diffusive system
including interferences {\hyperlink{Lubatsch05}{\cite{Lubatsch05}}$^,${\hyperlink{Frank11}{\cite{Frank11}}} that
incorporates non-linear effects and gain. All types of light-matter
interactions depend not only on the material and the passive refractive index
as well as the mobility of electrons, but further on the locally impinging
light intensity, the photon number. It has to be pointed out, that the
refractive index of the scatterers has to be renormalized selfconsistently due to intense pumping. This is
      equivalent to a shift of the gain spectrum with respect to impinging
      intensity taking into account in this case that the threshold of a
      random laser is defined as the stationary
      state. Consequentially we treat second order non-linear response of the
      bulk material when the order is defined in the electromagnetic field
      $E$. It has been seen in the previous section, that gain processes
      $-{\rm Im}\epsilon_s$ lead to a retardation of coherent intensity due to
      a finite life-time of the electronic excitation. $\epsilon_s$ is the
      permittivity of the scatterer. Frequency conversion, in other words
      spectral loss or gain, leads to a change of the photon statistics
      respectively. The refractive index of the material is responding to
      these processes and especially it is responding due to spatially
      non-uniform or non-homogeneous procedures. These procedures are present
      in every system containing any boundary, meaning in any realistic setup.\\

Theoretically the non-linearity is established
by a doubly nested self-consistency: In the following we line out the description for correlation and coherence of light in
terms of the electromagnetic wave and the photon as particle.\\
The photon density response, the four-point correlator is derived from
the Bethe-Salpeter equation of photons,

\begin{eqnarray}
\Phi = G^RG^A[1 + \int\frac{d^3q}{(2\pi)^3} \gamma\,\,\,\Phi]\label{BS1}
\end{eqnarray}
that is in the most essential form written including six independent position
coordinates

\begin{eqnarray}
\label{BSREAL}
\Phi (r_1,r_1';r_2,r_2')=&& G^R (r_1,r_1')G^A
(r_2,r_2')\,+\nonumber\\
&& \sum_{r_3,r_4,r_5,r_6}\!\!\!\!\!\!\!\!G^R (r_1,r_5) G^A
(r_2,r_6)\,\times\gamma(r_5,r_3;r_6,r_4) \Phi(r_3,r_1';r_4,r_2').
\end{eqnarray}
\\
The indices mark independent positions within space. Dashes denote the
selfconsistency procedure of the diagram. The irreducible vertex {\bf$\gamma$} in Eq.(\ref{BS1}) is the heart of the formalism. It contains all
possible interference effects enhanced by maximally crossed diagrams
  (Cooperons) which lead to a sophisticated current relaxation kernel, the memory term that renormalizes the diffusion
  constant $D$. We emphasize here that we assume independent monodisperse
  scatterers here and the Cooperon diagram is the leading order diagram under
  ensemble average where physically relevant system size is larger than the wavelength of light. The importance of the Cooperon has been clearly proven from our
  considerations of Anderson localizing systems in transmission with qualitative as well as
  quantitative validity {\hyperlink{Anderson}{\cite{Anderson}}}.
The memory
  kernel is crucial for the random laser mode. It establishes spatial
  correlation and coherence whereas the temporal coherence of lasing emission
  is driven by the interplay between these transport processes and the
  non-linear response of the material. The Ward identity as such is the	vital
  feature in photonic transport and in diagrammatic theory of two-particle
  propagators in general. It connects the single-particle
  Feynman-graph with the two-particle quantity {\hyperlink{Tiggelen}{\cite{Tiggelen}}}$^,${\hyperlink{PingSheng}{\cite{PingSheng}}} in conserving
  media ($Im \epsilon_s = Im \epsilon_b = 0$), and in recent work
  {\hyperlink{Lubatsch05}{\cite{Lubatsch05}}} it is generalized to guarantee local energy conservation, or specifically energy non-conservation for complex matter.\\
We write the Bethe-Salpeter equation as Boltzmann- or kinetic equation
Eq.(\ref{four}). The Fourier transformation and the expansion into momenta
yields the exact continuity equation for the correlator $\Phi$ with spatial
dependencies due to the loss channels at the boundaries of the finite system
and additionally the current density relation.

\begin{eqnarray}
\label{four}
&&[\Delta \Sigma + 2 {\rm {Re}}\,\epsilon_b \omega\Omega - 2
{\rm{Im}}\,\epsilon_b\omega^2 - 2\vec{p}_{x}\cdot \vec{Q}_{X}\, +\, 2ip_y\partial_Y]\,\, \Phi_{pp'}^{Q_{X}} (Y  , Y') \\\nonumber
&&= \Delta G_p (Q_{X}; Y, Y')\delta(p-p')+\sum_{Y_{3,4}} \Delta G_p (Q_{X})\,\times\,\int \!\!\!\frac{{\rm d}p'' }{(2\pi)^2}
\gamma_{pp''}^{Q_{X}}  (Y,Y_{3,4})
\Phi_{p''p'}^{Q_{X}}  (Y_{3,4} , Y'). \\\nonumber
\end{eqnarray}

$\Delta G = G^R -G^A$, $p$, $p'$ and $p''$ are momenta. The scatterer's
geometric properties are represented within the self-consistent complex valued
scattering matrices {\bf$T$} of the Schwinger-Dyson 
  equation {\bf$G=G_0 + G_0TG$} which leads to the solution for the Green's
  function {\bf$G^R$} and
  {\bf$G^A$} of the
  electromagnetic field, the light wave. The ZnO scatterer's initial permittivity is
  given by $Re\,{\epsilon_s=4.0164}$, the imaginary part $Im\,{\epsilon_s}$, the
  microscopic gain, is computed self-consistently yielding gain
  saturation. The photon density emitted from the amplifying Mie particles is
  derived by means of coupling to the rate equation system (see previous section). It is self-consistently connected the
dielectric function {\bf$\epsilon=\epsilon_L + \epsilon_{NL}$}. Finally we
arrive at nonlinear feedback in both, electromagnetic wave transport and
photon intensity transport for scalar waves. The scalar approach is especially
suitable to model absolutely randomized particle systems. Further the Mie
character develops with reducing particle size into a Rayleigh
scatterer and strong non-isotropies which might influence the vector-character
are consequentially not given. Only in setups of
pronounced Mie type scatterers as well as order or quasi-order we expect the
vector-character of light to become important. Random
lasing in such setups has not been investigated theoretically as well as experimentally so far.

Within grand canonical (open) ensembles of random lasers the entropy is
increased by photonic intensity transport processes. Nevertheless transport
in the meaning in hand is based on the time reversal symmetry of the single
particle Green's function $G = [\epsilon_b(\omega/c)^2 -|{\bf q}|^2 -
\Sigma^\omega_{\bf q}]^{-1}$ describing the propagation of the electromagnetic
wave. This time reversal symmetry is diagrammatically not broken. We describe the laser dynamics within a laser rate equation system that
is suitable for non-linear processes or quantum cascades {\hyperlink{Siegman}{\cite{Siegman}}}. The advantage is obvious since nonradiative decay processes within
that system act directly on the electronic subsystem, the particle, and so
enter directly the non-linear complex refractive index and the self-energy $\Sigma$ of
the single independent Mie scatterer, modelled as the complex scattering or {\bf T} matrix. For clarity it
shall be pointed out that the complex refractive index acts equally absorbing
or emitting under time reversibility. Microcanonically the time evolution is
flipped, however the system evolves grand-canonically {\it open}
{\hyperlink{Frank06}{\cite{Frank06}}}.

The described procedure of modeling disorder and dissipation guarantees the
completeness of the  'ab initio' description of the
propagating light intensity by the four-point correlator {\bf$\Phi =
  A\,\Phi_{\epsilon\epsilon} + B\,\Phi_{J\epsilon}$} here given in terms of the
momenta. {\bf$\Phi_{\epsilon\epsilon}$} equals the energy density and
{\bf$\Phi_{J\epsilon}$} equals the energy current, {\bf$A$} and {\bf$B$} are
pre-factor terms derived in {\hyperlink{Frank11}{\cite{Frank11}}}. The framework yields all
specific transport characteristics, e.g. the scattering mean free path {\bf$l_s$} and includes all interference effects. The mode is described efficiently by the determination of
the correlation length {\bf$\xi$} with respect to various loss
channels. This length {\bf$\xi$} in non-linear systems marks a
decay of the intensity to $1/e$. It is of a qualitative different importance
than the localization length in the Anderson sense {\hyperlink{Frank11}{\cite{Frank11}}}, because the
diffusion constant is $D\neq0$ in complex media. In other
words, the state in this case has a finite lifetime compared to the immanent
infinite lifetime of an Anderson localized state in a passive system. The Bethe-Salpeter equation is solved in a sophisticated regime of real space and momentum and the description
for the energy density {\bf$\Phi_{\epsilon\epsilon}(Q,\Omega)$} is derived which is
computed regarding energy conservation

\begin{eqnarray}
\label{Pole}
\Phi_{\epsilon\epsilon}(Q,\Omega)= \frac
{N_{\omega}(Y)}
{
\Omega + i D Q_{X}^2
\underbrace{- i D
  {\chi_{d}^{-2}}-{c_{1}\Big(\partial^{2}_{Y}\Phi_{\epsilon\epsilon}(Q,\Omega)\Big)+ c_{2}}   + i D  \zeta^{-2}}  _{i D \xi^{-2}} }.
\end{eqnarray}

The numerator {\bf$N_\omega$} is the local density of
photonic modes LDOS renormalized due to amplification and absorption of the
electromagnetic wave. $Q$ equals
the center of mass momentum of the propagator denoted in Wigner coordinates,
$\Omega$ is the center of mass frequency and $D$ is the self-consistently
derived diffusion constant. $c_1$, $c_2$ are coefficients having a non-trivial
form of only numerical relevance. Following the formal analytical structure of
Ref. {\hyperlink{Frank11}{\cite{Frank11}}} up to the result of Eq. (28) under the additional assumption of boundaries as well as
the coupling to laser rate equations (Eq.(\ref{lrgtot})) leads actually to some more sophisticated term. This result has to be reformed algebraically until it fits
in it's structure a formal diffusion pole again. The above
mentioned form of the energy density $\Phi_{\epsilon\epsilon}$ (Eq.(\ref{Pole}))
emerges. A divergence, that is marked by critical scales in the unlimited system of
{\hyperlink{Frank11}{\cite{Frank11}}}, is here instead replaced by the phase transition towards
lasing. Consequentially modes derived as the characterising result here equal not only coherent transport of photons
but they rather equal lasing modes caused by the inverted electronic sub-system in quasi-equilibrium, the stationary state. \\
The structure of the Bethe-Salpeter equation and the diffusion pole will be discussed in the following. The excitation process is uniform in space. Interferences gain weight on long paths in-plane of the large scaled
random laser sample. The physics of maximally crossed diagrams therefore significantly dominates
the coherence properties: Dissipation and losses due to
spontaneous emission and
non-radiative decay are in principle homogeneous, however they depend of course
very well to the impinging energy density and the resulting non-linear
response. As consequence these properties change with the position relative to
the samples boundaries if the latter are lossy. All channels are represented within the pole of Eq.(\ref{Pole}) resulting in
separate dissipative length scales $ \zeta $ due to homogeneous losses, and
$\chi_{d}$ due to gain and absorption that go along with photonic transport
and the open or strongly absorbing boundaries. All dissipation processes
enter the mass term of the diffusion equation:

\begin{eqnarray}
\!\!\!\!i D \xi^{-2}=
 - i D
  {\chi_{d}^{-2}}-{c_{1}\Big(\partial^{2}_{Y}\Phi_{\epsilon\epsilon}(Q,\Omega)\Big)+
    c_{2}}   + i D  \zeta^{-2}.
\label{diff}
\end{eqnarray}

By solving of the non-classical diffusion equation
Eq.(\ref{diff}) the coefficients $c_1$ and $c_2$ are selfconsistently
derived. Non-classical is defined as to consider light, as explained above,
diagrammatically not just as a wave but in addition as particle (photon)
current. Finally, we derive the spatial distribution of
energy density:

\begin{eqnarray}
-\frac{\partial^2}{\partial Y^2}  \Phi_{\epsilon\epsilon}
=
 \frac{1}{D}\left[
\frac{D}{- \chi_d^2} +\frac{D}{\zeta^2 }
\right]
\Phi_{\epsilon\epsilon}
+ {\rm ASE}.
\label{SE}
\end{eqnarray}

The nonlinear self-consistent microscopic random laser gain $\gamma_{21}
n_2$ incorporates the influences of both length scales $\chi_d$ and $\zeta$,

\begin{center}
\begin{eqnarray}
\frac{D}{- \chi_d^2 }+\frac{D}{\zeta^2 }
=
\gamma_{21}  n_2 ,
\end{eqnarray}
\end{center}

and therefore represents the physical properties of the random laser samples
within the absorptive boundaries. $\gamma_{21}$ is the transition rate of
stimulated emission and $n_2$ equals the selfconsistent occupation of the
upper laser level. The abbreviation {\bf$\rm ASE$} on the right of
Eq.(\ref{SE}) represents all transport terms yielding amplified spontaneous
emission. 

\section*{Results and Discussion}
\label{results_and_discussion}
\renewcommand{\baselinestretch}{2}
In the previous sections we developed a model for transport in strongly
scattering, dense particle agglomerates. Non-linear gain and gain saturation is included in the
model {\hyperlink{LRG}{Fig. \ref{LRG}}} by the coupling to a rate equation system, describing the full lasing
dynamics Eq.(\ref{lrgtot}).\\
In {\hyperlink{DUAL}{Fig. \ref{DUAL}}} we display the onsight on a lasing sample derived by calculations for co-existing random lasing modes
of ZnO powder on GaAs substrate. The co-existence is explained by a sharp
spectral separation of both mode types. Extended modes may only be developed
due to absorption at the samples boundaries as it has been shown in
{\hyperlink{Frank14}{\cite{Frank14}}}.  In these calculations the non-radiative
decay $\gamma_{nr}$ is assumed to be 0, so the sample features only losses at
the boundaries towards the substrate and out of plane in $z-$direction. Lasing
 principle occurs in all D=3 dimensions but lasing intensity escapes the
 sample only in the $z-$direction. Under the additional premise that the
observed lasing frequency is not absorbed at the samples boundaries in-plane
($x-y-$plane), which means the sample size is infinite compared to the mean
free path $l_s$, confined modes arise {\hyperlink{DUAL}{Fig. \ref{DUAL} (a)}}. Our result is
experimentally confirmed in samples of ZnO on SiO and GaAs {\hyperlink{Kalt}{\cite{Kalt}}}.

We are tuning now the phonon-production rate in order to investigate thermal
loss as an adjustment process for random laser modes. This is equivalent with
a microscopic detuning of the electronic subsystem, in other words a modified
quantum efficiency. Phonon production as a homogeneous rated process in the
whole system in first instance reduces of course spontaneous emission and
therefore rises the laser threshold as such. As a second process that is
subtle but even more important, it reduces coherent scattering and
interference effects. Long range interferences, represented in the Cooperon
diagram, however, trigger stimulated processes on large scales. If they are
reduced, not only the amount of coherently emitted intensity is smaller, also
the modes diameters generally are small {\hyperlink{Corr}{Fig. \ref{Corr}}}. The sample is changing
it's quality from the laser, which features a phase transition at the
threshold, towards a so called super-radiating system that has the
functionality of a large area light-emitting diode (LED). In Eq.(\ref{SE}) the
last term responsible for amplified spontaneous emission (ASE) gradually is increased.
The numerical analysis at the laser threshold can be found in
{\hyperlink{DUAL}{Fig. \ref{DUAL}(a)}} and {\hyperlink{Corr}{Fig. \ref{Corr}}}. For the confined modes it is found, that they behave
non-linear when the phonon-production is modified. In {\hyperlink{Corr}{Fig. \ref{Corr}}} the
diameter of the mode with varying $\gamma_{NR}$ is displayed. $\gamma_{NR}$ is
running over $0.0 .. 1.4$ in units of the spontaneous transition rate.
The dots on the curve mark equidistant steps of $0.2$. It is found in the deviation from the red line that the decrease of the diameter
is non-linearly behaving with the loss. Also the photon emission rate
shows a non-linearity but with an opposite impact. The distance between the dots for increasing $\gamma_{NR}$ is decreasing. This can be understood in
the picture of the Cooperon as the stimulating process for lasing
radiation as explained above. \\
Both modal regimes are not just
simply varying in their diameter, they are in our theory rather different from
the fundamental point of view. It can be excluded that non-radiative tuning will cause another modal
regime like an extended mode covering the whole sample as displayed in {\hyperlink{DUAL}{Fig. \ref{DUAL}\,(b)}}. This transition is not to be explained as an intrinsic
Stokes-shift. It is further noted that the extended mode, which is connected
by loss to the surrounding substrate suffers through additional non-radiative
losses in intensity. The amount of coherently emitted intensity is reduced. However their mode diameters are almost insensitive to
that loss type because they are pinned to the bondaries.\\

\section*{Summary and Conclusion}
\label{conclusion}
\renewcommand{\baselinestretch}{2}
We have shown in this work co-existing extended and confined random laser
modes. Extended modes occur according to our results definitely due to boundary absorption.
Tuning the quantum efficiency of large samples of random lasers by means of non-radiative decay
leads to a modulation of the lasing spot size of the modes in the confined
regime. However, in our framework we derive a non-linear dependency and a
decrease of the spot diameters with an increasing phononic action. The
transition from confined to extended modes, so the drastic increase of the
mode diameter, is
certainly not reachable by temperature or phonon production. Additionally we
have found in our theoretical analysis evidence that the Cooperon contribution
is reduced by absorption and supported by gain procedures even though the
Anderson transition in it's original sense is not given in open random
media. However it will be extremely interesting to understand these
fundamental procedures and to find out numerically in detail, how
the interplay of diffusion and localization procedures is responding to
gain. Up to our knowledge such a study has not been performed yet.\\

\section*{Acknowledgments}
\renewcommand{\baselinestretch}{2}
RF thanks H. Cao, H. T\"ureci and B. Altshuler for highly valuable discussions
on the change of random
laser modes due to heat or noise at the ICTP conference
{\it Coherent Phenomena in Disordered Optical Systems}.
\\
\\
\\

\vspace*{1.0cm}

{\bf Author contribution statement:} A. L. and R. F. wrote the manuscript and
prepared the figures 1-4. Both authors reviewed the manuscript.\\

{\bf Competing financial interests statement:} There are no competeing
financial interests.\\

\begin{figure}[b]
{\hypertarget{exp}{\color{white}Figure1}}
\begin{center}\resizebox{0.5\textwidth}{!}{%
\includegraphics[clip]{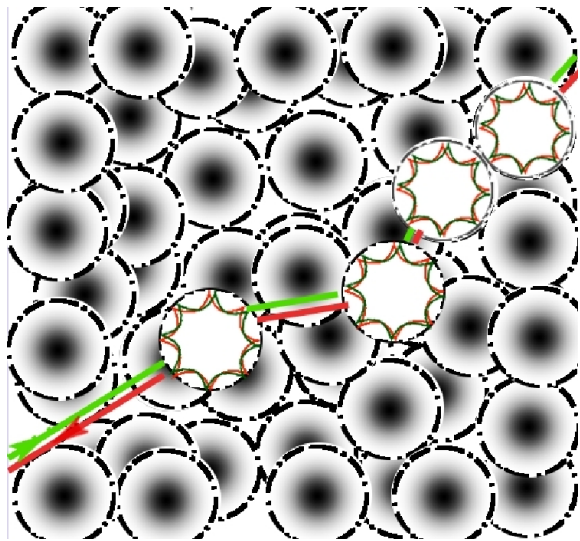}%
}\end{center}
\caption{Sketch of experimentally relevant sample. Displayed is a powder of
  monodisperse ZnO grains, diameter of the grains is $d\,=\,260nm$. The
  volume filling of the sample is assumed to be about $50\%$. The sample shall
  be thin in the direction of the incident pump beam but infinite
  in-plane. The red and green paths mark the propagation direction of
  scattered photons (red) and their time-reversal symmetric processes
  (green). The full correlation includes all interferences in the
  theory. Within the spheres photons experience amplification through
  extended light-matter bound states as well Mie resonances. They can be seen as whispering
  gallery modes propagating inside
      the sphere and being reflected at the surface determined by the
      refractive index contrast, which enhance
  forward-scattering \cite{Mie}.}
\label{exp}     
\end{figure}

\begin{figure}[b]
{\hypertarget{LRG}{\color{white}Figure2}}
\resizebox{0.5\textwidth}{!}{\includegraphics[clip]{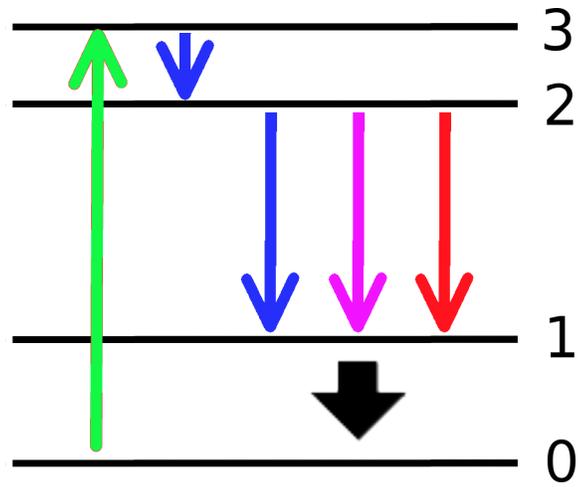}
}
\caption{Sketch of a 4-level laser rate system. Sinuous lines represent
  electronic procedures due to 2-photon pumping, the
exciton (green), spontaneous emission (blue), stimulated emission (bright
pink) and phonons (red).}
\label{LRG}     
\end{figure}

\begin{figure}[b]
{\hypertarget{DUAL}{\color{white}Figure3}}
\quad\hspace{0cm}\rotatebox{0}{\scalebox{1.0}{\includegraphics[width=1.0\textwidth]{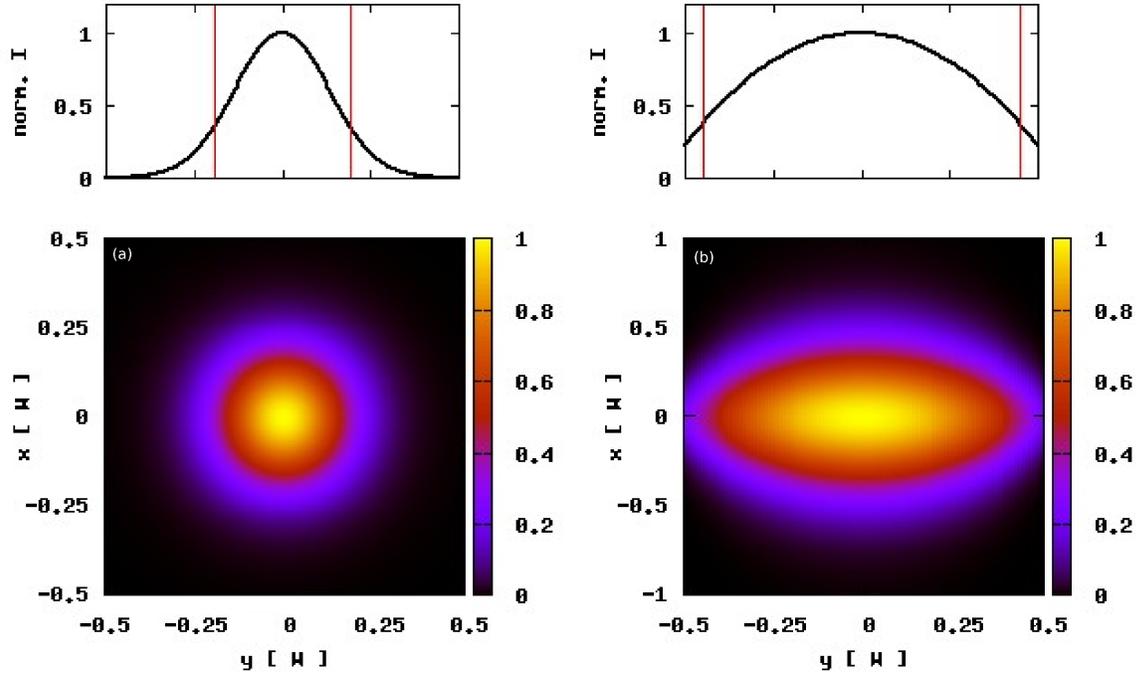}}}
\vspace*{0.0cm}\\
\caption{Computed coherent lasing intensity
      distribution (color bar). The ZnO scatterers, diameter $d=260\,
nm$, filling $50\,\%$, are embedded in a lossy
substrate i.e. GaAs or SiO. The dimensions are $W_x\,=\,20.0\mu m$,
$W_y\,=\,40.0\mu m\,++$, $W_z\,=\,4.0\mu m$. Results are shown for
  homogeneous 2-photon pumping $ \lambda = 355$ nm (bandedge of ZnO bulk). (a) Confined mode. Emission energy is
$3.23 eV$, the transport mean free path $l_s= 499.2$ nm. Shown is the result onsight on the samples section of $20.0\mu\, m \times \,20.0\mu\, m$, the mode features no
absorption at the samples edges. However the underlying substrate is
absorbing. (b) Extended mode. Sample section
$20.0\mu m\,\times\,40.0\mu m$. The profiles above the color coded plots show the
normalized coherent intensity $I$ of each mode. The difference due
to lossy boundaries in (b) is evident by comparing the decay to $1/e$ (vertical lines in (a)
and (b)). The phonon-production in this calculation is $\gamma_{NR} = 0$.}
\label{DUAL}
\end{figure}

\begin{figure}[b]
{\hypertarget{Corr}{\color{white}Figure4}}
\hspace{-3.0cm}\resizebox{0.8\textwidth}{!}{\includegraphics[clip]{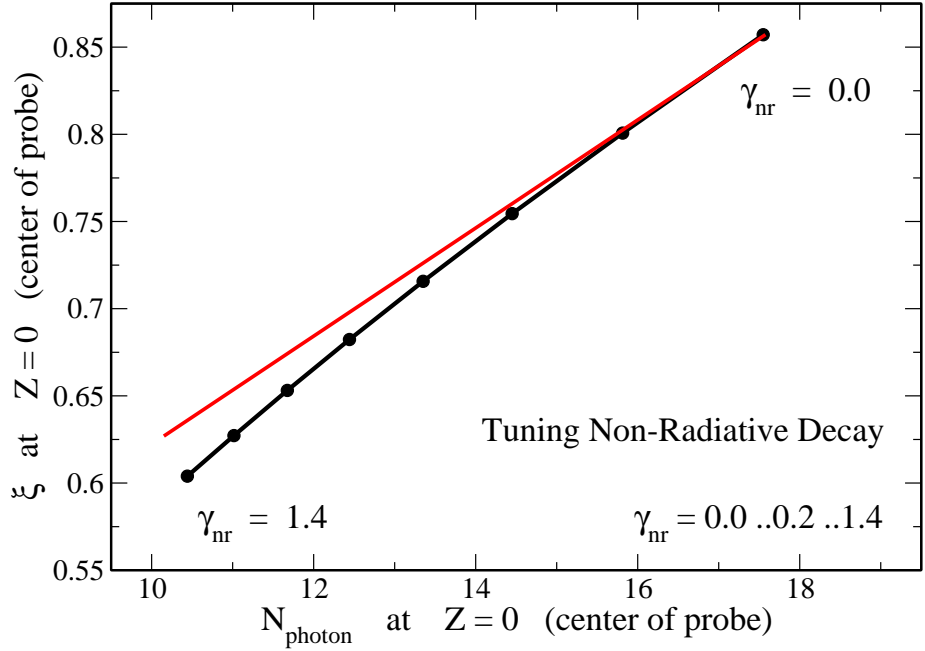}}
\caption{The computed correlation length $\xi$, the mode diameter of a
      bulk mode at the surface, as a function of the emitted photon number in the center of a confined
  mode which is tuned by non-radiative decay. The decay rate $\gamma_{NR}$ varys between $\gamma_{NR}\,=\,0.0$ and $\gamma_{NR}\,=\,1.4$ in units of the spontaneous
emission rate. The non-linear behavior (indicated by deviation from a
    linear behavior given as red line) is clearly enhanced dependent to strong
losses and the mode diameter is reduced about $30\%$. It is evident that the
mode is continuously decreased in the confined regime.}
\label{Corr}     
\end{figure}

\end{document}